\documentclass[runningheads]{llncs}
\usepackage[T1]{fontenc}
\usepackage{graphicx,verbatim}

\usepackage{hyperref}
\hypersetup{
  hidelinks
}
\usepackage{color}
\usepackage{amsmath, amsfonts, amssymb}
\usepackage{float}
\usepackage{booktabs}
\usepackage{subfig}
\usepackage{enumitem}
\usepackage{bbm}
\usepackage{gensymb}
\usepackage{mathtools}
\usepackage{multirow}
\usepackage{multicol}
\usepackage{xspace}
\usepackage{svg}

\makeatletter
\DeclareRobustCommand\onedot{\futurelet\@let@token\@onedot}
\def\@onedot{\ifx\@let@token.\else.\null\fi\xspace}
 
\def\ie{\emph{i.e}\onedot}

\newlist{inlist}{enumerate*}{1}
\setlist[inlist]{label=\arabic*)}
\newlist{inlistalpha}{enumerate*}{1}
\setlist[inlistalpha]{label=\alph*)} %

\begin{document}
\title{Non-intrusive Body Composition Assessment from Full-body mmWave Scans}

\author{Miriam Senne\inst{1,2\ast}\orcidID{0009-0008-6995-3033} \and
Benjamin~D.~Killeen\inst{1,2\ast}\orcidID{0000-0003-2511-7929}\and \\
Tony Danjun Wang\inst{1,2}\orcidID{0000-0002-0883-1744}
\and
Nassir Navab \inst{1,2}}
\authorrunning{M. Senne et al.}
\institute{Chair for Computer Aided Medical Procedures\\
Technical University of Munich, Boltzmannstr. 3, 85748 Garching, Germany
\email{\{miriam.senne,bd.killeen,tony.wang,nassir.navab\}@tum.de}
\and
Munich Center for Machine Learning, Munich, Germany\\
\noindent \small $^{\ast}$Equal contribution
}

\maketitle              %
\begin{abstract}
Body composition assessment (BCA) provides detailed information about the distribution of different tissue types in the body, enabling more precise characterization of individuals than BMI or weight alone.
Consistent and frequent BCA would be valuable for personalized medicine, but the gold standard methods for BCA, such as CT and MRI, are only practical for opportunistic monitoring of patients with clinical indications for imaging and are not suitable for routine use in the general population.
Here, we consider an imaging modality which is not currently used in medical applications: millimeter wave (mmWave) radar. Commonly used in security settings, mmWave scans enable fast, non-intrusive, and privacy-preserving reconstruction of full body shape without the need to remove clothing.
To demonstrate the feasibility of fast and convenient BCA from mmWave scans, we present a method for BCA value regression using a multi-task learning strategy that leverages synthetic mmWave-like point clouds derived from clinical imaging and parametric human models. We evaluate the model on a pilot cohort of real mmWave scans with bioimpedance-derived body fat measurements, supporting the feasibility of estimating VAT and body fat percentage (BFP) from mmWave data acquired through clothing in a standing posture. We find that the model can predict VAT and BFP with a mean absolute error of 1.0 L and 3.2\%, respectively, demonstrating the potential of mmWave scanning for routine BCA in a wide range of settings.

\end{abstract}
\section{Introduction}

Body composition assessment (BCA) is an important component in the emerging landscape of personalized medicine, providing detailed information about the distribution of different tissue types in the body. Quantitative measurements of skeletal muscle volume, abdominal fat volume, or bone mineral density, for example, characterize individuals more precisely than body mass index (BMI) or weight, and studies have repeatedly demonstrated their clinical utility for risk stratification and prognosis in a wide range of conditions, including cardiometabolic disease~\cite{quirino2025cardiometabolic,pickhardt2021opportunistic}, cancer cachexia (\ie, rapid muscle degeneration), and frailty in elderly patients~\cite{bates2022ct}. Consistent and frequent BCA would expand these and other benefits to the general population. At the same time, with the maturity of fully automated algorithms for medical image analysis~\cite{wasserthal2023totalsegmentator}, and in particular for segmentation of CT and MRI scans, there is growing interest in opportunistic body composition assessment as a byproduct of routine clinical imaging~\cite{killeen2025towards}.  Recent work has shown that external body shape alone can provide strong signals for adiposity and fat distribution, including models that predict body fat percentage from three dimensional body point clouds~\cite{zheng2024d3bt} and surface meshes~\cite{mueller2023body}, as well as models that estimate VAT and related depots from two dimensional silhouettes derived from MRI~\cite{klarqvist2022silhouette}. Here, we consider an imaging modality which is not currently used in medical applications: millimeter wave (mmWave) radar. Commonly used in security settings, mmWave scans enable fast, non-intrusive, and privacy-preserving reconstruction of full body shape without the need to remove clothing~\cite{anonymous}. It can be deployed in a wide range of settings, including primary care clinics and community centers. Our goal is to enable fast and convenient BCA from routine mmWave scans, with a particular focus on the estimation of visceral adipose tissue (VAT) and body fat percentage (BFP).
\begin{figure}[t]
\centering
\includegraphics[width=\linewidth]{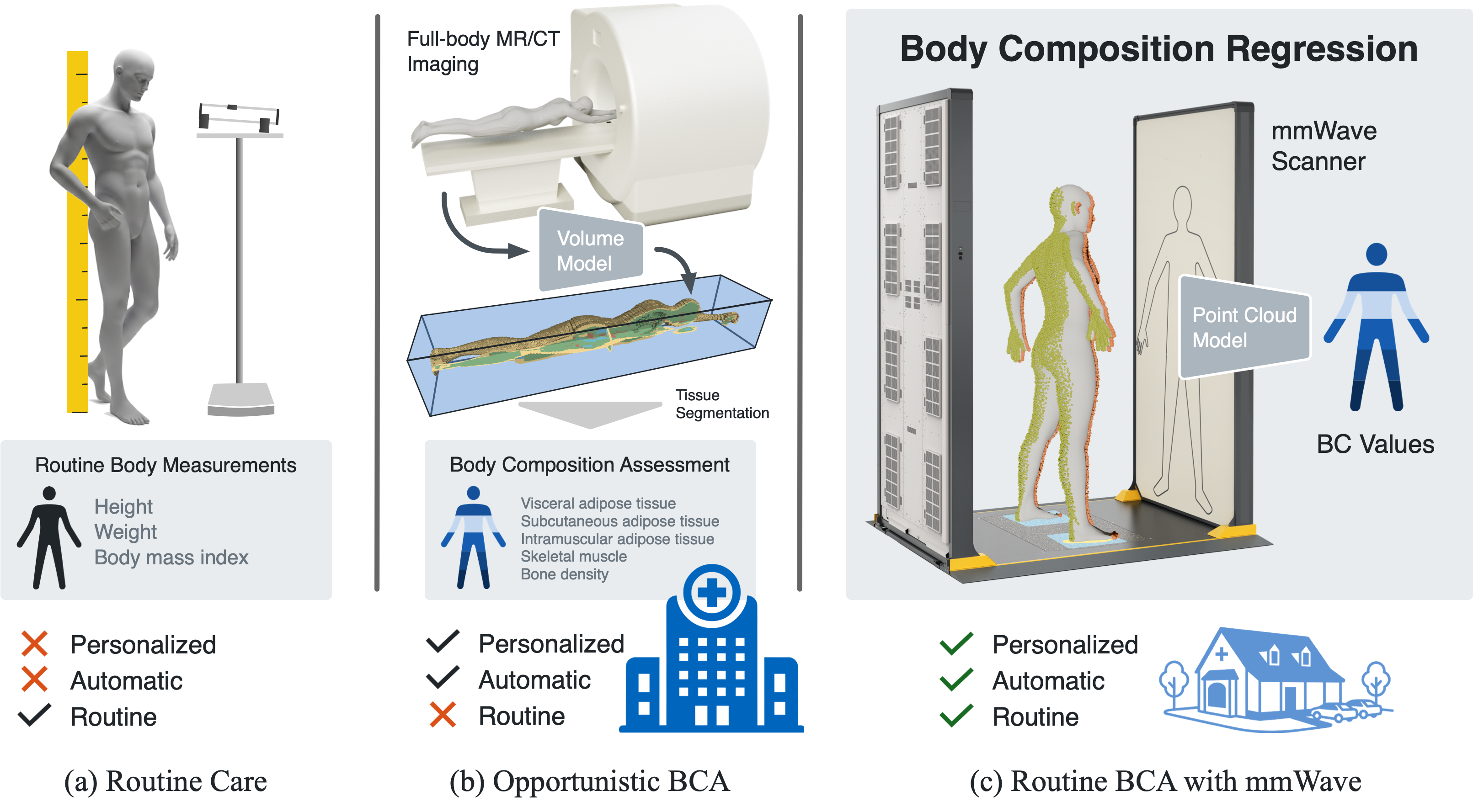}
\caption{Routine body measurements like height, weight, and BMI are widely used but provide limited information about body composition (a). In contrast, detailed body composition assessment (BCA) can provide more precise characterization of individuals and has demonstrated clinical utility, but is currently only practical for opportunistic monitoring of patients with clinical indications for imaging (b). Millimeter wave (mmWave) radar is a fast, non-intrusive, and privacy-preserving imaging modality that can reconstruct a partial body shape without the need to remove clothing, enabling BCA value regression (c).}
\label{fig:overview}
\end{figure}

In this paper, we present a method for VAT and BFP regression from mmWave scans. After reconstruction, the output of an mmWave scan is a point cloud representing the front and back surface of the body, with missing data in the regions with a surface normal not aligned with the AP axis. To train a model that estimates internal body composition from this surface geometry, we confront two main challenges:
\begin{inlist}
\item the lack of paired mmWave and body composition data, and
\item the domain gap between supine or prone MRI/CT scans and standing mmWave scans.
\end{inlist}
Using tissue type segmentation~\cite{wasserthal2023totalsegmentator} of MRI and CT scans is a fast, scalable way to obtain VAT and BPT values, and we introduce a method to synthesize mmWave-like point clouds from these scans based on the body surface geometry. However, because these scans are acquired in the supine or prone position, the observed surface geometry is affected by gravitational deformation and contact with the support surface on the anterior and posterior sides, respectively. Alternative sources of body surface data, like the Skinned Multi-Person Linear Model (SMPL)~\cite{loper2023smpl}, can provide a complete body surface in arbitrary poses, but lack any information about the internal body composition. The same is true of available datasets of mmWave scans, although other measurements can be automatically derived from body surfaces~\cite{anonymous} as well as CT/MRI scans~\cite{killeen2025towards}. Thus, we propose a multi-task learning strategy that leverages each data source for the information it provides, with the goal of learning a shared representation that can bridge the domain gap between the different modalities and postures. Specifically, we train a single point cloud encoder with multiple regression heads, regressing volumes of intramuscular adipose tissue (IMVAT), visceral adipose tissue (VAT), subcutaneous adipose tissue (SAT), total body, lean tissue (LT) and muscle (MV); height; and chest, waist, and hip circumferences, using posterior and anterior surface data from MRI/CT scans, SMPL meshes, and mmWave scans. We demonstrate the feasibility of this approach by evaluating the model on a withheld set of synthetic mmWave-like point clouds derived from MRI/CT scans and SMPL models, the real mmWave dataset including anthropometric measurements, and on a pilot cohort of real mmWave scans with bioimpedance-derived body fat measurements.

\section{Methods}

\subsection{Dense Point Cloud Extraction from CT/MRI Data}
Paired millimeter wave point clouds and ground truth body composition labels are expensive to acquire and are therefore scarce. To enable supervised training, publicly available full body MRI and CT datasets with body composition annotations are used. From each volume, surface geometry is extracted and converted into point clouds that simulate mmWave scanner output. Because the scanner provides full body coverage, only full body reference scans are considered. Full body MRI scans are provided in the HIT dataset \cite{keller2024hit}, where subjects were scanned in the prone position stretching their arms over their heads, leading to detailed information of the back surface while producing a limited and flat representation of the anterior surface.
The HIT dataset contains 398 whole body MRI scans with four tissue classes, namely muscle, SAT, IMVAT, and bone and  fitted SMPL body models. IMVAT corresponds to a combined segmentation of VAT and intramuscular adipose tissue (IMAT). A binary body volume is constructed by retaining background as label 0 and mapping all other tissue labels with values greater than 0 to 1. Since air cavities such as the lungs may be labeled as background, slice based morphological operations are applied to obtain a closed and solid body volume. A surface mesh is then reconstructed using marching cubes with a threshold of 0.5.
To obtain complementary anterior surface information, we employ a subset of full body CT scans from the NMDID \cite{edgar2020new} database, which contain examinations of diseased subjects that have their arms crossed in front of their chest.
The NMDID subset used in this work is filtered to include only cases of natural death with minimal decomposition, resulting in 798 full body CT scans. An additional 41 scans are excluded due to extreme BF below 5\% or above 70\%. Tissue labels are generated using TotalSegmentator, and the labels for SAT, VAT, and skeletal muscle are retained. Surface meshes are reconstructed from the body label, which covers the trunk and the remaining body regions, using marching cubes.
To obtain a high resolution and smooth surface suitable for point cloud synthesis, Laplacian smoothing is applied to all reconstructed meshes, after which 800000 points and corresponding normals are sampled from each surface. Body composition labels are computed as tissue volumes by counting voxels assigned to the corresponding tissue class and multiplying by the voxel volume derived from the image spacing provided with each scan. The resulting values are converted to liters. 

\subsection{Dense Point Cloud Reconstruction from Parametric Models}

To bridge the domain gap between supine/prone clinical imaging and the standing posture required for mmWave scanning, we extend our training dataset with synthetic point clouds derived from the SMPL model \cite{loper2023smpl}.
We establish base postures—specifically, an upright standing position with arms slightly abducted (A-pose) and a posture with arms crossed in front of the chest, resembling CT acquisitions (X-pose).
We captured informal reference photographs of these poses and fit SMPL parameters to them \cite{SMPL-X:2019}.
For each base pose, we generate a variety of plausible human shapes by sampling the SMPL shape parameters ($\beta$) with a standard deviation that ensures a realistic, diverse distribution of body measurements and overall composition.

Achieving high pose diversity alongside this shape variation presents a distinct challenge, as naively applying noise to pose parameters often results in implausible articulations, such as unnaturally bent joints.
To overcome this, we perturb the initial poses and project them back onto the closest plausible configuration on a learned manifold of valid human articulations using Neural Riemannian Distance Fields (NRDF) \cite{he2024nrdf}.
Finally, we employ VolumetricSMPL \cite{mihajlovic2025volumetricsmpl} to guarantee that the resulting meshes are physically realistic and free of self-intersections, which we then convert into synthetic mmWave point clouds.
Since the HIT dataset provides fitted SMPL models that exclude the flattened regions caused by the scan bed, these meshes are used directly for the prone poses. SMPL meshes in A-pose and X-pose are sampled in the same quantity as the HIT SMPL instances, yielding a total of 1194 SMPL based samples. Resulting meshes are Laplacian smoothed and 800000 surface points with corresponding normals are sampled.
Anthropometric labels are derived by measuring chest, waist, and hip circumferences according to ISO 8559 \cite{ISO8559}. For each shaped SMPL mesh, the model is first placed in a canonical pose. Axial cross sections are then extracted at fixed keypoint locations, and each circumference is computed as the perimeter of the convex hull of the corresponding cross section.

\subsection{Data Pre-Processing and Normal Filtering}
Microwave imaging is derived from radar and exploits electromagnetic waves in the 30 to 300 GHz range to penetrate materials and interact with internal structures \cite{ahmed2012advanced}. The used mmWave sensor comprises two opposing panels, and subjects stand between them in an A pose as shown in Fig. \ref{fig:overview}. Acquisition requires 25 ms to 30 ms and yields an intensity volume with 1.9 mm by 1.9 mm lateral resolution and 5.5 mm depth resolution, where high values correspond to the skin surface. Due to viewpoint dependent illumination, regions with surface normals oriented away from the panels do not get captured, resulting in missing scanner observation, particularly along the body sides. Since front and back volumes are available separately, point clouds are extracted by locating depthwise intensity maxima and applying thresholding. Synthesized mmWave data from meshes have to exhibit those properties.

Prior to training, each point cloud is registered to the millimeter wave scanner coordinate space. Augmentation is applied using a random rotation about the vertical axis, with angles sampled from a normal distribution and clipped to the range from -10 to 10 degrees. Positional jitter is added with a maximum magnitude of 2 mm to reflect the lateral scanner resolution.
A normal based filtering step is then used to simulate scanner illumination. For each point, the squared dot product between the point normal and a panel normal vector is computed, and points below a threshold are removed. The threshold is sampled from a normal distribution with mean 0.7 and is treated as an additional augmentation parameter, as well as a subsequent random down-sampling to 100000 points.

\subsection{Network Architecture}

\begin{figure}[t]
\centering
\includegraphics[width=0.85\linewidth]{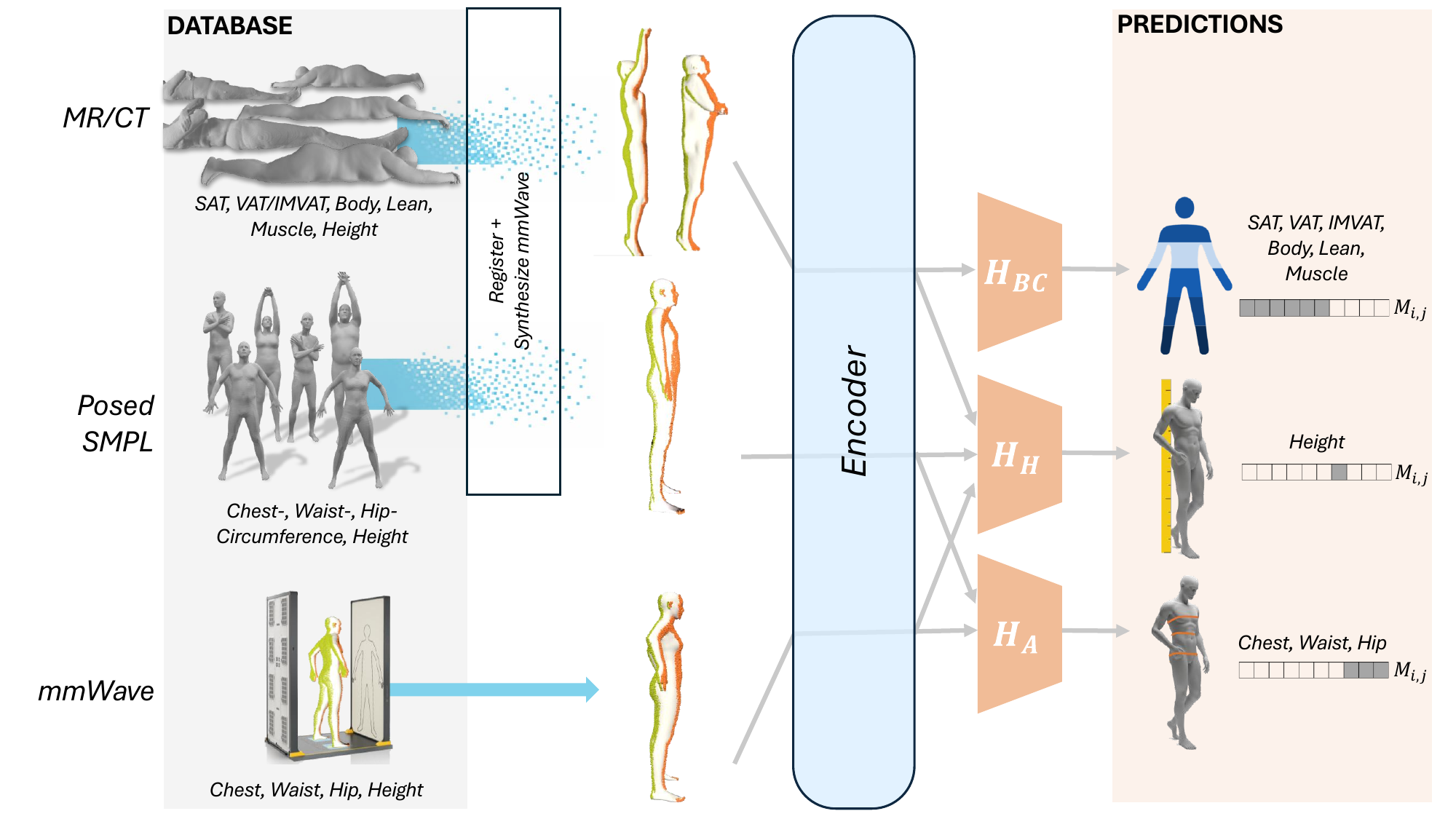}
\caption{Overview of the proposed pipeline, including mmWave specific point cloud synthesis from CT, MR, and SMPL surfaces, for a $n$-d multi-headed model with a shared encoder and three task-specific regression heads.}
\label{fig:data-model}
\end{figure}

The proposed multi-headed regression model, shown in Fig.~\ref{fig:data-model}, predicts ten body related measurements from three dimensional point clouds. The backbone is a PointTransformerV3 encoder \cite{wu2024point} initialized from large-scale pretraining on diverse indoor three dimensional datasets, since pretrained weights for human point clouds are not available. Since the pretrained configuration expects six input feature channels, the three dimensional coordinates are duplicated to form an N by 6 feature tensor. The encoder extracts a shared 512-dimensional feature representation from three dimensional point clouds, which is passed to three task specific MLP heads. ${Head}_H$ predicts body height and outputs one value, while ${Head}_A$ predicts three anthropometric measurements and ${Head}_{BC}$ predicts six body composition metrics.
Each head uses the same MLP design. The input layer receives the 512-dimensional encoder feature, followed by two hidden layers with 256 and 128 units. Each hidden layer is followed by layer normalization, GELU activation, and dropout. A final linear layer maps the 128 dimensional representation to the required output dimension.

All target labels are Z-score normalized using a precomputed mean and standard deviation for each target. In multi task learning, manual selection of loss weights is challenging because tasks differ in scale and convergence behavior. This work therefore uses a masked and dynamically weighted objective that supports partial labeling and stabilizes optimization. First, we define a localized Masked Sample Loss $(\ell_{i,h})$ to ensure gradient isolation:

\begin{equation}
\ell_{i,h} = \frac{\sum_{j \in \text{Head}_h} \mathcal{L}(\hat{y}_{i,j}, y_{i,j}) \cdot M_{i,j}}{\left( \sum_{j \in \text{Head}_h} M_{i,j} \right) + \epsilon}
\label{eq:masked_loss}
\tag{2}
\end{equation}

The binary mask $M_{i,j}$ zeros out missing labels, while the denominator normalizes the gradients. This allows the shared encoder to learn universal geometric features from disjoint datasets. The per target loss $\mathcal{L}$ is the Huber loss, which is robust to outliers via a quadratic to linear transition. The small constant $\epsilon$ ensures stability when no labels are present for a given head. Second, we employ Dynamic Weight Averaging (DWA) to adaptively balance the $H$ heads:

\begin{equation}
\mathcal{L}_{total} = \sum_{h=1}^{H} w_h(t) \cdot \left[ \frac{1}{N_h} \sum_{i=1}^{N} \ell_{i,h} \right]
\label{eq:total_dwa_loss}
\tag{3}
\end{equation}
By monitoring relative descent rates, DWA \cite{liu2019end} automatically prioritizes lagging tasks. The constraint $\sum w_h=H$ stabilizes the total gradient energy, ensuring robust optimization without manual hyperparameter tuning.

Training uses AdamW with separate learning rates for the encoder (0.00005) and heads (0.0001), a cosine annealing scheduler, and a 10-epoch warmup for stable optimization. Weight decay and gradient clipping are applied for regularization and stability. On a single Nvidia RTX 6000 Ada GPU, training for 150 epochs with a batch size of six takes about six hours.

\section{Evaluation}

\newcommand{\NA}{\multicolumn{1}{c}{---}}
\begin{table}[t]
\centering
\caption{MAE and standard deviation on withheld test sets, reported per dataset for the pretrained multi-headed model. *BFP in \% is included based on dataset specific fat definition.}
\scriptsize
\begin{tabular}{@{}llrrrrrrrrrrr@{}}
\toprule
&& Height (cm) & \multicolumn{3}{c}{Circumference (cm)} & \multicolumn{4}{c}{Volume (L)} & \multicolumn{1}{c}{BFP$^\ast$ (\%)} \\
\cmidrule(lr){4-6} \cmidrule(lr){7-10}
~~~ & Dataset & & \multicolumn{1}{c}{Chest} & \multicolumn{1}{c}{Waist} & \multicolumn{1}{c}{Hip} & \multicolumn{1}{c}{SAT} & \multicolumn{1}{c}{IMVAT} & \multicolumn{1}{c}{VAT} & \multicolumn{1}{c}{Body} & \\
\midrule
\multicolumn{2}{l}{\textit{Synthetic}}\\
& NMDID & $1.9_{\pm 1.6}$ & \NA & \NA & \NA & $2.5_{\pm 2.2}$ & \NA & $1.0_{\pm 0.9}$ & $2.0_{\pm 1.5}$ & $4.07_{\pm 3.30}$ \\
& HIT & $1.6_{\pm 1.3}$ & \NA & \NA & \NA & $2.2_{\pm 1.5}$ & $0.7_{\pm 0.7}$ & \NA & $1.6_{\pm 1.1}$ & $2.36_{\pm 1.54}$ \\
& $\text{SMPL}_{\sum}$ & $1.2_{\pm 0.9}$ & $1.2_{\pm 1.0}$ & $1.5_{\pm 1.2}$ & $1.1_{\pm 0.8}$ & \NA & \NA & \NA & \NA  & \NA\\
\multicolumn{2}{l}{\textit{Real}}\\
& $\text{mmWave}_{A}$ & $1.1_{\pm 0.4}$ & $2.8_{\pm 1.3}$ & $1.8_{\pm 1.3}$ & $1.5_{\pm 0.6}$ & \NA & \NA & \NA & \NA  & \NA \\
& $\text{mmWave}_{BC}$ & $1.4_{\pm 1.0}$ & \NA & \NA & \NA & \NA & \NA & \NA & \NA & \NA  \\
\bottomrule
\end{tabular}
\label{tab:results}
\end{table}

Evaluation is conducted on synthetic point clouds derived from CT, MR, and SMPL data, as well as on real millimeter wave scans. For the HIT dataset, the provided train/val/test split of 8/1/1 is used. All remaining datasets are split across subjects and gender with the same ratio. In addition to the synthetic data, we use a real mmWave anthropometry dataset~\cite{anonymous}, denoted $mmWave_{A}$, with 54 scans from 27 subjects.
Table~\ref{tab:results} shows the performance of the model on withheld test sets for each dataset, using mean absolute error (MAE), where $\text{SMPL}_{\sum}$ denotes aggregated results across all SMPL subsets. The results show that across synthetic datasets, anthropometric errors are typically within 1 to 2\,cm, indicating that the model compensates for the mmWave illumination gaps.

\begin{figure}[t]
\centering
\begin{minipage}{0.48\textwidth}
\centering
\includegraphics[width=\linewidth]{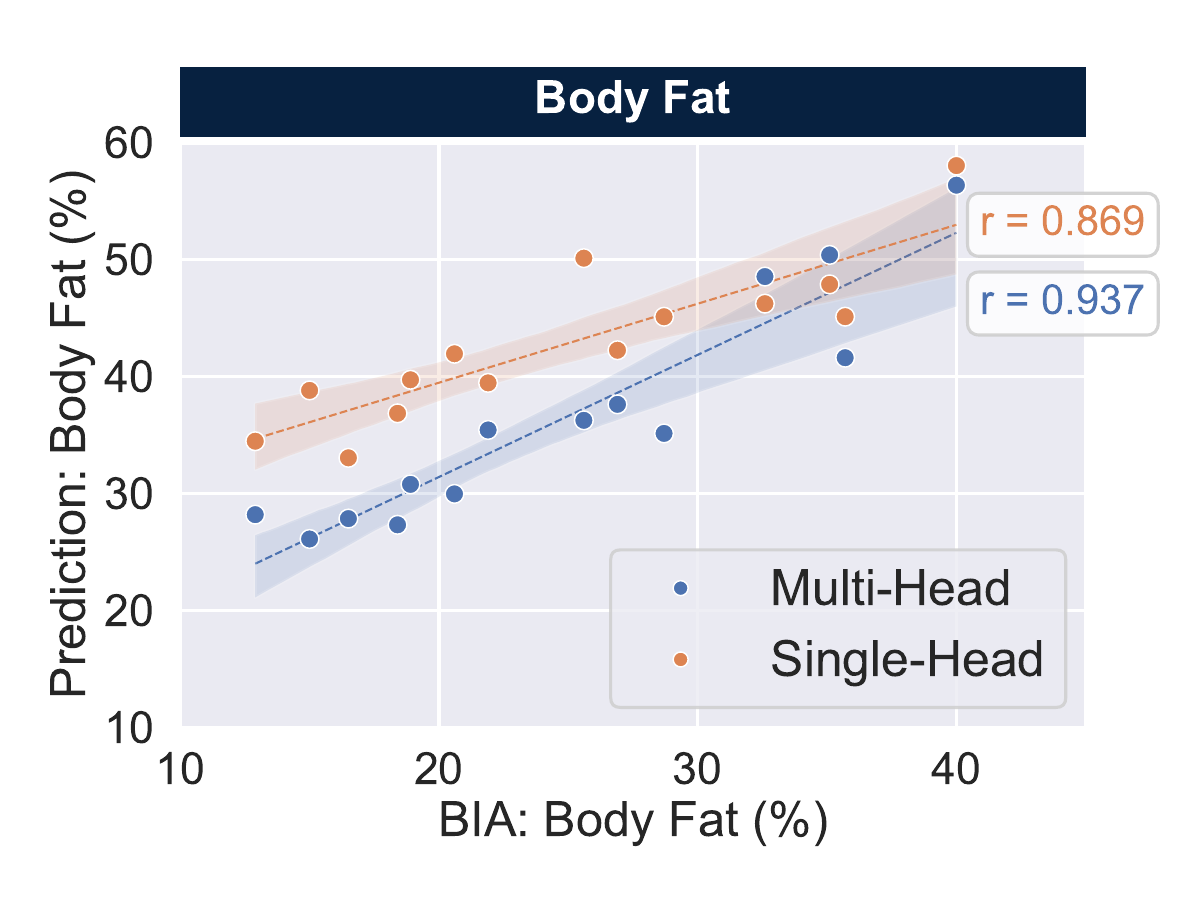}
\end{minipage}
\hfill
\begin{minipage}{0.48\textwidth}
\centering
\includegraphics[width=\linewidth]{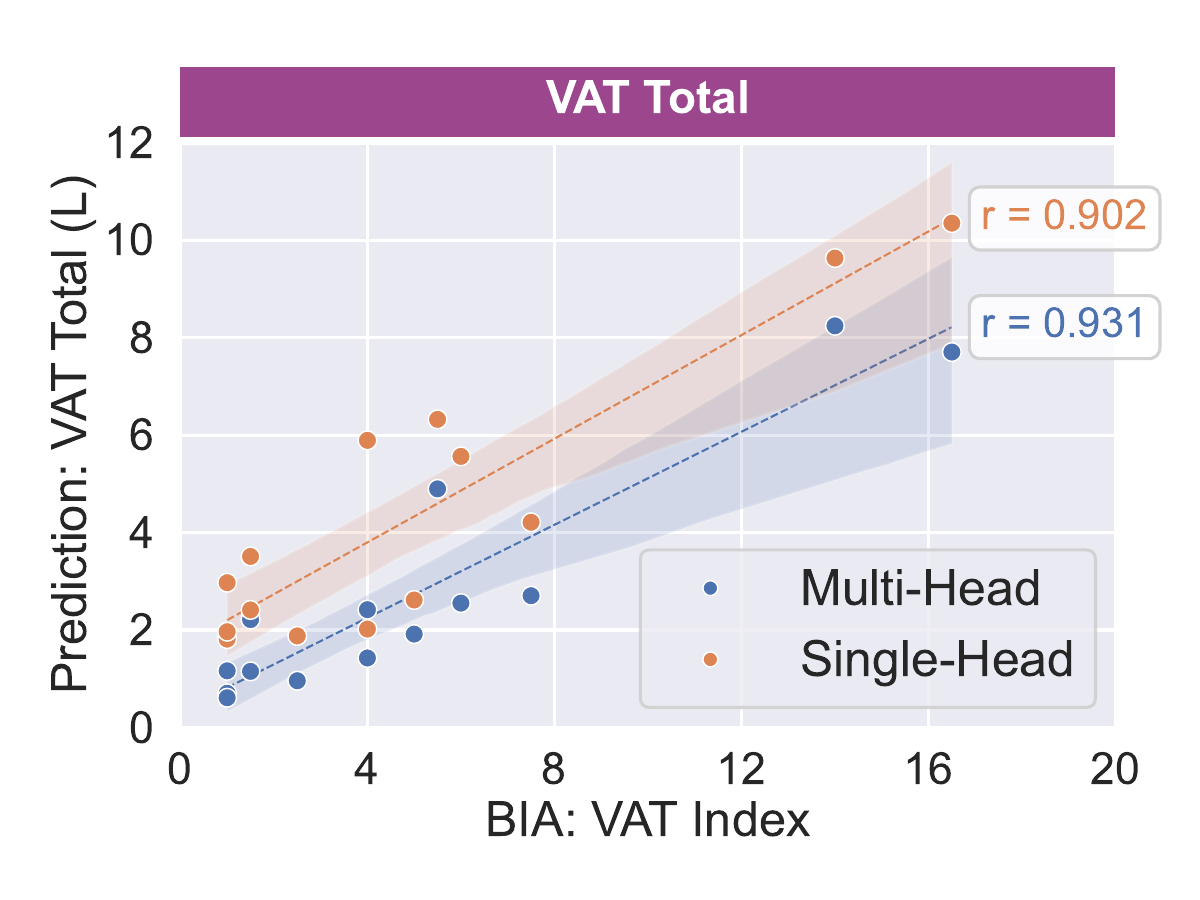}
\end{minipage}

\caption{Correlation of BFP (left) and VAT (right) with a 95\% confidence interval for the $\text{mmWave}_{BC}$ data, for multi-head and single-head architectures. Pearson score $r$ rates the correlation of predicted and target values.}
\label{fig:results}
\end{figure}

To assess transfer to real mmWave scans for BCA, a pilot cohort dataset, denoted $\text{mmWave}_{BC}$, was collected and comprises scans of 14 subjects with paired height and BCA values acquired using the Tanita RD 545HR bioelectrical impedance scale \cite{TanitaRD545HR}. Participants signed an informed consent and ethics waiver was granted. To reduce variability in impedance based estimates, measurements were performed in the morning and participants were instructed to avoid food and drink beforehand. Model outputs are compared against Tanita VAT and BFP measurements. The model predicts VAT volume in liters, whereas Tanita reports a VAT index in the range 0 to 59, where values from 1 to 12 indicate a healthy level and values above 12 indicate an elevated level. BFP is used as a second metric, and a comparable value is computed as the sum of IMVAT, or VAT when IMVAT is unavailable, and SAT divided by the total body volume.
For completeness, this derived BFP MAE is also reported in Table \ref{tab:results} for the synthetic CT/MR data. Because the VAT index has no physical unit conversion and the derived BFP may be affected by error propagation from multiple predicted volumes, agreement is assessed using correlation.
Associations are visualized in Fig. \ref{fig:results} and quantified using the Pearson correlation coefficient $r$. We also evaluate a single head $H_{BC}$, only trained on CT and MR body composition data, to evaluate the effect of multi task learning. The multi-headed model shows consistently stronger association than the single head baseline, with $r$ increasing from 0.868 to 0.937 for BFP and from 0.902 to 0.931 for VAT, indicating that jointly learning related biomarkers improves transfer to real mmWave scans.

\section{Discussion and Conclusion}

Routine assessment of body composition has the potential to enhance risk stratification and support preventive care, but current methods are limited by cost, accessibility, and patient burden. mmWave scanning offers a promising alternative, providing rapid, non-invasive, and clothing-penetrating measurements that can be performed in a standing position without undressing. In this paper, we demonstrated the feasibility of mmWave-based BCA using multi-task regression model with a shared encoder trained on synthetic point clouds derived from CT, MR, and SMPL surfaces, as well as real mmWave scans. Further work is needed to characterize the model's performance with paired mmWave-MRI datasets, rather than commercial bio-impedance measurement devices. Although these devices offer a potentially even more convenient alternative for BCA than mmWave scanners, they are prone to systematic error and do not measure detailed variations of fatty tissue~\cite{tewari2018a}. Despite these limitations, the model successfully predicted anthropometric measurements and body composition metrics, showing strong correlation with reference standards in the pilot cohort.

\section{Acknowledgment}
This work has been supported by Rohde \& Schwarz. Additionally, we thank the TUM Ethics Committee for their advice regarding data collection. All participants provided informed consent.

\bibliographystyle{splncs04}
\bibliography{main}

\end{document}